\documentclass[10pt, final, journal, letterpaper, twocolumn]{IEEEtran}
\makeatletter
\def\ps@headings{%
\def\@oddhead{\mbox{}\scriptsize\rightmark \hfil \thepage}%
\def\@evenhead{\scriptsize\thepage \hfil \leftmark\mbox{}}%
\def\@oddfoot{}%
\def\@evenfoot{}}
\makeatother 

\IEEEoverridecommandlockouts

\usepackage{amsfonts}
\usepackage[dvips]{graphicx}
\usepackage{times}
\usepackage{cite}
\usepackage{amsmath}
\usepackage{array}
\usepackage{amssymb}

\newcommand{\bs}{\boldsymbol}

\usepackage{stfloats}
\usepackage{slashbox}
\usepackage{graphicx}
\usepackage{footnote}
\usepackage{booktabs}
\usepackage{array}
\usepackage{algorithmic}
\usepackage{algorithm}
\usepackage{subeqnarray}
\usepackage{cases}
\usepackage{threeparttable}
\usepackage{color}

\begin{document}

\title{Energy Harvesting for Physical-Layer Security in OFDMA Networks}

\author{Meng Zhang and Yuan Liu,~\IEEEmembership{Member,~IEEE}
\thanks{The authors are with School of Electronic and Information Engineering,
South China University of Technology, Guangzhou, 510641, P. R. China. Email: akjihfdkog@gmail.com, eeyliu@scut.edu.cn.
}
}
\maketitle

\vspace{-1.5cm}
\begin{abstract}
In this paper, we study the simultaneous wireless information and power transfer (SWIPT) in downlink multiuser orthogonal frequency-division multiple access (OFDMA) systems, where each user applies power splitting scheme to coordinate the energy harvesting and secrecy information decoding processes. Assuming equal power allocation across subcarriers, we formulate the problem to maximize the aggregate harvested power of all users while satisfying secrecy rate requirement of individual user by joint subcarrier allocation and optimal power splitting ratio selection. Due to the NP-hardness of the problem, we propose two suboptimal algorithms to solve the problem in the dual domain. The first one is an iterative algorithm that optimizes subcarrier allocation and power splitting in an alternating way. The second algorithm is based on a two-step approach that solves the subcarrier allocation and power splitting sequentially. The numerical
results show that the proposed methods outperform conventional methods. It is also shown that the iterative algorithm performs close to the upper bound and the step-wise algorithm provides good tradeoffs between performance and complexity.

\end{abstract}


\section{Introduction}
Orthogonal frequency division multiplexing access (OFDMA) gains its popularity and has become a leading multiple access candidate scheme for beyond fourth/fifth generation (4G/5G) wireless systems,
due to its flexibility in resource allocation and robustness against multipath fading. It enables efficient transmission of various data
traffic by optimizing power, subcarrier, and bit allocation among different users. 
%

Due to the broadcasting nature of wireless channels, security is a crucial issue in designing wireless communication systems. As a traditional method, cryptography encryption
dominates the upper layers mainly by increasing the  complexity  has been introduced in every
layer but physical layer in the standard five-layered protocol stack. Thus physical-layer security is an important
complement to the other security approaches.

A great deal of studies have been devoted to the information-theoretic physical-layer
security\cite{Shannon,Wyner,Leung,Jorswieck,Li,Goel,OFDMADF,Wang,Kwan}. For example,
in \cite{Jorswieck,Li}, resource allocation for physical-layer security considerations was studied for multicarrier systems. In \cite{Goel}, artificial noise was considered for physical-layer security.
%
However, the artificial noise based methods for physical-layer security mainly lie on the spatial degrees of freedom offered by multiple antennas to degrade the channel of the eavesdroppers.
%
Alternatively, simultaneous wireless information and power transfer (SWIPT) also becomes an important solution to improve the energy utilization for wireless networks by prolonging the
lifetime of wireless nodes and draws a great deal of research interests \cite{Fouladgar,ZhangHo,ZhouZhang,LiuPS}. Energy harvesting wireless networks are potentially able to gain energy from the wireless environments. The prior work \cite{Fouladgar} studied the performance of SWIPT in the receiver that can decode information and harvest energy for the same received signal, which may be not realizable however. Two practical schemes, so-called time switching and power splitting, were proposed
in \cite{ZhangHo,ZhouZhang} as practical designs. With time switching applied at a receiver, the received signal is either
processed for energy harvesting or for information decoding. When the power splitting is applied at the receiver, the received signal
can be split into two streams with one stream processed by the energy receiver and the other processed by the information
receiver. The authors studied the flat-fading channel variations in SWIPT in \cite{LiuPS}, where dynamic power splitting was applied in the systems. Two SWIPT schemes in OFDMA with different configurations and corresponding resource allocation problems were studied in \cite{HuangLarsson}.


A handful of works have studied SWIPT for physical-layer security, usually considering some receivers decode confidential information and the rest receivers harvest energy (also known as the \textit{separated} receiver model) \cite{LiuGC13,fading,robustSWIPT}.
These works are mainly motivated by the dual use of the artificial noise, i.e., artificial noise is used to
interfere with the eavesdropper for secrecy information receivers and acts as the source of energy harvesting for energy receivers.
 Such method is efficient for the separated receivers, however not for the \textit{co-located} receivers where the receivers can simultaneously receive secrecy information and harvesting energy.
%
%
Greatly different from the existing solutions for physical-layer security (such as artificial noise and beamforming), we consider a \textit{co-located} SWIPT system by using power splitting scheme which is an ``SWIPT" way against eavesdropping in an OFDMA system. Specifically, if subcarriers are preferable to transmit secrecy information, the user may split more received power for information decoding and, on the contrary, if subcarriers are easily eavesdropped, the user splits more received power for energy harvesting, which helps the systems to fully utilize both spectrum and energy of the easily eavesdropped subcarriers that are traditionally difficult to utilize.

%

One challenge is that power splitting should be performed before OFDM demodulation in practical OFDMA-based SWIPT systems. Thus each user should split the power of all received information on all subcarriers at a same ratio instead of dynamic ratio. The complexity of designing power splitting ratio on all subcarriers with the same ratio is much greater than that of splitting on each subcarrier with dynamic
ratio, because in the former case, the power splitting ratio of each user couples all subcarriers in the rate expression.

In this study, we consider the secrecy-rate required  downlink multiuser OFDMA networks, where all users apply power splitting scheme (of a same power splitting ratio at all subcarriers) to coordinate energy harvesting and information decoding processes. By assuming equal power allocation at subcarriers, our goal is to maximize the aggregate harvested power of all users while satisfying the secrecy rate constraint of each user by jointly optimizing the subcarrier allocation and designing the power splitting ratio.

We formulate the problem with power splitting applied at each receiver for practical application (P-PA)  as  a mixed integer programming problem and NP-hard. Since the optimal solution is difficult to obtain, we introduce two suboptimal algorithms with polynomial time complexity. We first propose an efficient iterative algorithm to find the power splitting ratio and subcarrier allocation in an alternating way. To further reduce the complexity, we also propose a two-step algorithm that first obtains the optimal subcarrier allocation policy and then solves the optimal power splitting ratio. It is shown to tradeoff the complexity and performance. Numerical results show that the proposed iterative algorithms perform close to the performance upper bound and both proposed algorithms outperform the heuristic methods.

%


%

The rest of this paper is organized as follows. In Section II, we formulate the problem. In Section III, we propose two solutions. We study the case of statistical CSI of eavesdropper in Section IV. In Section V, the performance of the two schemes are evaluated via numerical results. Finally, we conclude with a brief summary of our results in Section VI.

\section{System Model And Problems Formulation}
In this paper, we consider a downlink OFDMA network, which consists of one base station (BS) with one antenna, $K$ mobile single-antenna users, over $N$ subcarriers and one single-antenna eavesdropper attempting to wiretap information from all subcarriers. 
It is trivial to extend to the non-cooperative multi-eavesdropper scenario, since the overall eavesdropped rate is the maximum rate of the multiple non-cooperative eavesdroppers. Thus the proposed algorithms are also applicable if we select the best eavesdropper link among multiple eavesdroppers on each subcarrier, i.e., the eavesdropper with the highest decodable information rate.
Each user communicates with BS and demands a secrecy rate that is no lower than a constant $C_{k}\geq0$, for all $1\leq k\leq K$. Here we assume
that equal power allocation is performed by the BS over all subcarriers for simplicity. This is reasonable since the gain brought
by power adaption is limited in OFDMA systems \cite{YuanTWC10,YuanTCOM12,YuanTWC13,YuanWCL12,TaoYuanTWC13}.
Each receiver is considered to split the received signal into two signal streams, with one stream to the energy receiver and the other one to information receiver.

The considered OFDMA-based SWIPT method for physical-layer security can be applied in various
scenarios, such as a home internet-of-things. In the considered example, the wireless
devices such as phones and tablets are simultaneously receiving confidential information and
harvesting energy from the wireless access point (like Wi-Fi or femtocells). The wireless access point
uses OFDMA to transmit signals (it is supportable in many standards). However, the neighbours
in/around the building attempt to eavesdrop the secrecy information.

We assume that all users are legitimate users and they have their own data transmission with the BS so that the BS can obtain full  channel state information (CSI) of users. Let $h_{k,n}$ denote the channel gain of user $k$ on subcarrier $n$, and $\beta_{n}$ denote the channel gain of the eavesdropper on subcarrier $n$. We also assume that each $\beta_{n}$ is independent and identically distributed (i.i.d.) Rayleigh fading channel.



Let $p_{n}$ represent the fixed and equal power allocated on subcarrier $n$. The received signal at user $k$ is processed by a power splitter, where we assume a ratio
$\rho_{k}$ of power is split to energy receiver and the remaining $1-\rho_{k}$ of power is split into the information receiver for OFDM demodulation. We have $0\leq \rho_{k} \leq 1$, $\forall k$. Note that power splitting is performed in analog domain before the
digital domain where OFDM demodulation is processed. Thus, due to this hardware limitation, each user has to harvest the received signal with a same power splitting ratio on all subcarriers.


{With the full CSI of eavesdropper known to the BS, the achievable secrecy rate at subcarrier $n$ of user $k$ is given by \cite{Shannon}
\begin{align}
r_{k,n}^{s}=&(r_{k,n}-r_{e,n})^+\nonumber\\
= & \left[ \log_2\left ( 1+\frac{\left ( 1-\rho _{k} \right )p_{n}h_{k,n}}{\sigma ^{2}} \right)
- \log_2\left(1+\frac{p_n\beta_n}{\sigma^2}\right) \right]^{+},\label{eqn:sr1}
\end{align}
where $[\cdot]^+=\max\{\cdot,0\}$, $\sigma ^{2}$, $r_{k,n}$ and $r_{e,n}$ are the power of additive white Gaussian noise, the achievable information rate of user $k$ and the eavesdropper, respectively. Note that the full CSI of eavesdropper case is practically valid in following scenarios: (i) the eavesdropper is active in the network so that the BS can monitor its behavior and obtain its CSI; (ii) interestingly, as stated in \cite{Ma2013}, even an passive eavesdropper's CSI can be obtained through its local oscillator power inadvertently leaked from the receiver RF front end using the methods in\cite{detecteve,detectCR}; (iii) the legitimate users and the eavesdropper belong to different networks in today's heterogeneous network, then the BS can coordinate with the  eavesdropper's serving network to obtain the CSI, since the eavesdropper is the legitimate user of different  network or service. This is referred to as coordinated multi-point (CoMP) transmission in 3GPP LTE-A. The assumption is widely adopted in the physical-layer security literature (e.g., \cite{Ma2013,Chen2014,Poor,Bloch,Zou2013}).

The secrecy rate of user $k$ is given by

\begin{equation}\label{eqn:pb1}
     r_{k}^{s}=\sum_{n=1}^{N}x_{k,n}r_{k,n}^{s},
\end{equation}
where we let $x_{k,n}$ denote the binary subcarrier allocation variable, with $x_{k,n}=1$ indicating that subcarrier
$n$ is assigned to user $k$ and $x_{k,n}=0$ otherwise. Note that if $\sum_{k=1}^{K}x_{k,n}=0$ for any subcarrier $n$, i.e., such subcarrier is not assigned to any user, then it is used to transmit power to users only.

With the conversion efficiency of the energy harvesting process at each receiver denotes as  $0<\zeta <1$, the
harvested power of user $k$ is thus given by
     \begin{equation}
     E_{k}=\zeta \rho _{k} \sum _{n=1}^{N}p_{n}h_{k,n}.
     \end{equation}

The goal of the considered problem is to find the optimal subcarrier allocation and power splitting ratio to maximize the total harvested power (for the purpose of uplink transmission for example) while satisfying the individual secrecy rate requirement for each user. This practical application optimization problem can thus be expressed as
     \begin{eqnarray}
 {\rm (P-PA):} ~~\max_{\{\bs X, \bs\rho\}}&&\zeta \sum _{k=1}^{K}\rho _{k} \sum _{n=1}^{N}p_{n}h_{k,n}\label{eqn:max}\\
{\rm s.t.}~&&\sum _{k=1}^{K}x_{k,n}\leq 1, \forall n \label{eqn:conx1}\\
  &&x_{k,n} \in \left \{ 0,1 \right \}, \forall k,n\label{eqn:conx2}\\
  &&0\leq \rho_{k} \leq 1,\forall k\label{eqn:conrho}\\
&&\sum _{n=1}^{N}x_{k,n}r_{k,n}^{s}\geq C_{k}, \forall k\label{eqn:conr} \end{eqnarray}
where $\bs X\triangleq\{x_{k,n}\}$ and $\bs \rho\triangleq\{\rho_{k}\}$.
The constraints in \eqref{eqn:conx1} and \eqref{eqn:conx2} enforce that each subcarrier can only be used by one user to avoid the multi-user interference.

\section{Proposed Algorithms}
The formulated (P-PA) is nonconvex due to the binary subcarrier variable $x_{k,n}$, finding the optimal solution is usually prohibitively due to the complexity.
However, according to \cite{noncon}, the duality gap becomes zero in multicarrier systems as the number of subcarriers
goes to large and the time-sharing condition is satisfied. Thus the optimal solution of a nonconvex resource allocation problem in multicarrier systems can be obtained in the dual domain.

Nevertheless, as we will discuss later, the traditional Lagrangian decomposition cannot be directly employed to decompose the problem into parallel subproblems with each subproblem corresponding to one subcarrier. This is because the power splitting ratio $\rho_k$ appears in the rate expression and couples the subcarrier assignment variables. As a result, solving (P-PA) is nontrivial though the dual method is used in this paper. In this section, we propose two efficient suboptimal algorithms.

 \subsection{Iterative Algorithm}

We define $ \mathcal{T} $ as all sets of possible $\bs X$ that satisfy \eqref{eqn:conx1} and \eqref{eqn:conx2}, $ \mathcal{R} $ as all sets of possible $\bs \rho$ that satisfy $0\leq\rho_{k}\leq1$.

 The Lagrangian function for (P-PA) is given by
  \begin{align}
 &L(\bs \rho ,\bs X,\bs \mu )\nonumber\\
 &=\sum_{k=1}^{K} \zeta \rho _{k}\sum _{n=1}^{N} p_{n}h_{k,n}+\sum _{k=1}^{K}\mu _{k}\left (\sum _{n=1}^{N} x_{k,n}r_{k,n}^{s}-C_{k} \right ) \nonumber\\
 &=\sum _{k=1}^{K}\sum _{n=1}^{N}\left (\zeta \rho _{k}p_{n}h_{k,n} +x_{k,n}\mu _{k}r_{k,n}^{s} \right )-\sum _{k=1}^{K}\mu _{k}C_{k},\label{eqn:lagfuc}
\end{align}
where $\bs \mu = \left[\mu_1,\mu_2,...,\mu_k\right]^{T}$ are the Lagrange multipliers. The dual function is then defined as
  \begin{equation}
g(\bs \mu)=\max_{\bs X \in \mathcal{T} ,\bs \rho \in \mathcal{R}}L(\bs \rho ,\bs X,\bs \mu )\label{eqn:dual2}.
\end{equation}

The dual problem is thus given by $\min_{\bs \mu}g(\bs \mu)$.
For the maximization problem in \eqref{eqn:dual2}, the Lagrangian function cannot be decomposed into $N$ subproblems, because the power splitting ratio $ \rho_k$ has to be computed considering all subcarriers that are assigned to user $k$, instead of one specific subcarrier.

Thus, for given dual variables $\bs\mu$, we can obtain a suboptimal solution by iteratively optimizing $\bs X$ with fixed  $\bs \rho$, and optimizing $\bs \rho$ with fixed $\bs X$. The process is repeated until both $\bs X$ and $\bs \rho$ converge, which is known as the block coordinate descent (BCD) method \cite{Iteration}.

To solve  $\bs X$ with fixed $\bs \rho$, suppose that subcarrier $n$ is assigned to user $k$, we have
\begin{equation}
L=\sum_{n=1}^{N}L_{n}-\sum_{k=1}^{K}\mu_k C_k,
\end{equation}
where
   \begin{equation}
  L_{n}=\zeta p_{n}\sum _{k=1}^{K}\rho _{k}h_{k,n}+\mu _{k}r_{k,n}^{s}. \label{eqn:solvex}\\
   \end{equation}
Thus, the subproblem is given by
   \begin{equation}
  \max_{\bs X_n \in \mathcal{T}}L_n(\bs \rho ,\bs X_n,\bs \mu )
   \end{equation}
which can be solved independently. By maximizing each $L_{n}$, the optimal $\bs X$ can be obtained as
   \begin{eqnarray}\label{eqn:opxpb2}
     x_{k,n}^{*}=\begin{cases} 1, ~{\rm if}~k=k^{*}=\arg \max_{k} ~L_n\\
     0, ~{\rm otherwise}.\end{cases}
   \end{eqnarray}

To solve $\bs \rho$ with  given $\bs X$, the problem can be decomposed into $K$ subproblems with each corresponding to one user since each $\rho_k$ is fixed in this process, which can be solved independently. The subproblem at user $k$ is given by
   \begin{eqnarray}
  \max_{\rho_k \in \mathcal{R}}L_{k}(\rho_{k})=\sum _{n=1}^{N}\left ( \zeta \rho _{k}p_{n}h_{k,n}+x_{k,n}\mu _{k}r_{k,n}^{s} \right ),\label{eqn:solverho}
   \end{eqnarray}
and we have
\begin{equation}
L=\sum_{k=1}^{K}L_{k}-\sum_{k=1}^{K}\mu_k C_k.
\end{equation}
Applying the Karush-Kuhn-Tucker (KKT) conditions, we have each $\rho_{k}^*$ has to satisfy
   \begin{align}
   &\frac{\partial L_{k}}{\partial \rho_{k}}=\sum_{n=1}^{N} \left [ \zeta p_{n}h_{k,n}-\frac{\mu_{k}x_{k,n} h_{k,n}p_{n}}{\ln 2\left ( h_{k,n}p_{n}\left ( 1-\rho _{k} \right)+\sigma ^{2} \right )} \right ]=0.\label{eqn:cannotsolve}
   \end{align}

 Unfortunately, there is no closed-form expression for the optimal $\rho_{k}^{*}$. However, since $L_k$ is a concave function of $\rho_{k}$, and $\frac{\partial L_{k}}{\partial \rho_{k}}$ monotonically decreases as $\rho _{k}$ increases, we can adopt the bisection search method to solve $\rho_{k}^*$ over $0\leq\rho_{k}\leq1$.

Nevertheless, an asymptotic solution can be obtained by considering a high received signal-to-noise (SNR) scenario, i.e., $\sigma^2\rightarrow0$. We have
   \begin{align}
&1-\rho_k=\frac{\mu_k}{\zeta\ln 2}\frac{\sum_{n=1}^Nx_{k,n}}{\sum_{n=1}^Np_{n}h_{k,n}}. \label{eqn:asl}
   \end{align}
In \eqref{eqn:asl}, $\frac{\mu_k}{\zeta\ln 2}$ is a constant in each iteration, $\sum_{n=1}^Np_{n}h_{k,n}$ is user $k$'s total received power which is also a constant, and $\sum_{n=1}^Nx_{k,n}$ is the number of subcarriers allocated to user $k$. Thus, we can conclude that $1-\rho_{k}$, the ratio of the power splitting into user $k$'s information receiver, is proportional to the number of subcarriers allocated to this user in high SNR scenario.

With the fixed $\bs \rho$, the optimal $\bs X^*$ can be obtained by \eqref{eqn:opxpb2}. The optimal value of the objective function can be increased by optimizing $\bs X$ via \eqref{eqn:cannotsolve}. Then, with the fixed $\bs X^*$, the optimal $\bs\rho^*$ can be obtained. Thus, the above process can be iterated until the optimal value of the objective function ceases to increase.

%

Finally, according to \cite{Boyd}, the dual function in \eqref{eqn:dual2} is always convex. By simultaneously updating $\bs \mu$, we can solve this problem by the subgradient method. The dual variables $\bs \mu$ are updated in parallel as
   \begin{align}
\mu_k^{(t+1)}=\left[\mu_k^{(t)}+\alpha_k\left ( C_{k}-\sum_{n=1}^{N}x_{k,n}r_{k,n}^{s}  \right)\right]^+,\forall k. \label{eqn:muupdate}
   \end{align}

   \begin{algorithm}[tb]
\caption{Proposed Iterative Algorithm for (P-PA)}
\begin{algorithmic}[1]
\STATE \textbf{initialize} $\bs \rho$ and $\bs {\mu}$.
\REPEAT
\REPEAT
\STATE Solve assignment variables $\bs X$ according to \eqref{eqn:opxpb2} and compute  $L$ according to \eqref{eqn:lagfuc}.
\FOR {each user $k$}
\STATE \textbf{initialize} $\rho_{k}^{UB}=1$ and $\rho_{k}^{LB}=0$.
\REPEAT
\STATE Set $\rho_{k}=\frac{1}{2}\left ( \rho_{k}^{UB}+\rho_{k}^{LB} \right )$.
\STATE Compute $\frac{\partial L_{k}}{\partial \rho_{k}}$ according to \eqref{eqn:cannotsolve}.
\IF {$\frac{\partial L_{k}}{\partial \rho_{k}}>0$}
\STATE Set $\rho_{k}^{LB}=\rho_{k}$.
\ELSE
\STATE Set $\rho_{k}^{UB}=\rho_{k}$.
\ENDIF
\UNTIL {$\left |\frac{\partial L_{k}}{\partial \rho_{k}} \right|<\varepsilon $, where $\varepsilon$ is a very small constant for controlling accuracy.}
\ENDFOR
\UNTIL {Lagrangian function converges.}
\STATE Update $\bs \mu$ by \eqref{eqn:muupdate} according to the ellipsoid method.
\UNTIL {$\bs \mu$ converge.}

\end{algorithmic}

\end{algorithm}

The above iterative algorithm to solve (P-PA) is summarized in Algorithm 1. For this algorithm, the complexity mainly lies in step 19). As each $\rho_{k}$ is obtained individually by the bisection search, the complexity of steps 6)-15) is $\mathcal{O}(K)$. Hence, The complexity of steps 4)-16) is given by $\mathcal{O}(K+KN)$. Next, the complexity of subgradient updates is polynomial in $K$ \cite{Boyd}. The overall complexity is given by $\mathcal{O}(K^{q+1}+K^{q+1}N)$, where $q$ is a constant and equal to $2$ for the ellipsoid method.

 \subsection{Step-Wise Algorithm}

Since the complexity of the above algorithm becomes unfavorable for practical application with the increase of $K$ and $N$, we also propose a simpler suboptimal algorithm in this subsection.

To begin with, we first formulate a problem by assuming that the power splitting can be designed differently on each subcarrier at each receiver. In this case, $\rho_k$ is extended to $\rho_{k,n}$, denoting the power splitting ratio on subcarrier $n$ at user $k$. Thus, we consider the following optimization problem as
     \begin{eqnarray}
 {\rm (P-UB):} ~~\max_{\{\bs X, \bs\rho\}}&&\zeta\sum _{k=1}^{K}  \sum _{n=1}^{N}\rho _{k,n}p_{n}h_{k,n}\label{eqn:maxuuper}\\
{\rm s.t.}~&&\sum _{k=1}^{K}x_{k,n}\leq 1, \forall n \\
  &&x_{k,n} \in \left \{ 0,1 \right \}, \forall k,n\\
  &&0\leq \rho_{k,n} \leq 1, \forall k,n\\
&&\sum _{n=1}^{N}x_{k,n}r_{k,n}^{s}\geq C_{k}, \forall k,
     \end{eqnarray}
where $r_{k,n}$ given in \eqref{eqn:sr1} is replaced by
\begin{align}
r_{k,n}=\log_2\left(1+\frac{(1-\rho_{k,n})p_{n}h_{k,n}}{\sigma^2}\right).
\end{align}

The optimal solution to this problem is given in Appendix A. As we have discussed in Section II, (P-UB) is hard implemented in currently practical receiver circuits. However, it gives a performance upper bound for the comparison purpose in simulation. Moreover, since (P-UB) can be directly decomposed into several subproblems (details in Appendix A), its solutions also provide useful insights  to design the step-wise algorithm due to its low complexity.

This step-wise algorithm is executed by two stages. The first stage is to seek the optimal subcarrier allocation policy $\bs X^*$ and the second stage is to find the optimal power splitting ratio $\bs\rho^*$. The two stages are separable instead of correlative as in the proposed iterative algorithm.

\begin{algorithm}[tb]
\caption{Proposed Step-wise Algorithm for (P-PA)}
\begin{algorithmic}[1]

\STATE Obtain $\bs X^*$ by solving (P-UB) according to Algorithm 3, given the same inputs (the same channel conditions and secrecy rate requirements).

\FOR {Each user $k$}
\STATE \textbf{initialize} $\rho_{k}^{UB}=1$ and $\rho_{k}^{LB}=0$.
\REPEAT
\STATE Set $\rho_{k}=\frac{1}{2}(\rho_{k}^{UB}+\rho_{k}^{LB})$.
\STATE Compute $r_{k,n}^{s}$ according to \eqref{eqn:sr1}.
\IF {$r_{k}^{s}>C_{k}$}
\STATE Set $\rho_{k}^{LB}=\rho_{k}$.
\ELSE
\STATE Set $\rho_{k}^{UB}=\rho_{k}$.
\ENDIF
 \UNTIL {$\left | r_{k}^{s}-C_{k} \right |<\varepsilon C_{k}$.}
 \ENDFOR

\end{algorithmic}

\end{algorithm}

The main idea of this algorithm is to first obtain the optimal subcarrier allocation variables $\bs X^*$ by solving (P-UB), then select power splitting ratio $ \bs \rho$ by the bisection search individually.
We first deduce the following theorem.

\textit{Theorem 1}: The optimal subcarrier allocation variables $\bs X^*$ for (P-UB) is also feasible for (P-PA), given the same inputs (same channel conditions and secrecy rate requirements).

\begin{IEEEproof}
Please see Appendix B.
\end{IEEEproof}

As a result, the proposed step-wise algorithm is feasible as long as the optimal algorithm for (P-UB) is feasible.

Moreover, for each user $k$, $r_{k,n}^s$ is monotonically decreasing in $\rho_{k}$. Therefore, $\bs \rho^*$ can be obtained by the bisection search.

The above algorithm is summarized in Algorithm 2. According to Appendix A, the complexity of solving (P-UB) is given by $\mathcal{O} ({K^{q}N})$. In addition,  $\rho_{k}$ is obtained individually by the bisection search, whose complexity is $\mathcal{O} ({K})$. Consequently, the complexity of the proposed step-wise algorithm is $\mathcal{O} ({K^{q+1}N})$ which is much lower than that of the proposed iterative algorithm.

\section{Case of Eavesdropper's Partial CSI}

In this section, we consider a more practical case where only statistical CSI of the eavesdropper is known at the BS, i.e., the BS only knows the CSI distribution of the eavesdropper. The CSI distribution of the eavesdropper can be acquired as follows in practice: assuming that the eavesdropper and legitimate users are randomly located in the cell (i.e., the eavesdropper and legitimate users
follow the same distribution), then the BS knows the eavesdroppers CSI distribution if the BS obtains the legitimate users CSI distribution.  Note that this assumption is widely used in the literature \cite{SecCapFad,FadingEve,Ergo1,Ergo2,Ergo3} and more practical than the previous case of full CSI of the eavesdropper.

The performance metric is the ergodic secrecy rate given by
\begin{align}
&r_{k,n}^{s}=(r_{k,n}-r_{e,n})^+\label{eqn:sr2}\\
= & \left\{ \log_2\left ( 1+\frac{\left ( 1-\rho _{k} \right )p_{n}h_{k,n}}{\sigma ^{2}} \right)
- \mathbb{E}_{\beta_n}\left[\log_2\left(1+\frac{p_n\beta_n}{\sigma^2}\right)\right] \right\}^{+}\nonumber\\
=&\left\{ \log_2\left ( 1+\frac{\left ( 1-\rho _{k} \right )p_{n}h_{k,n}}{\sigma ^{2}} \right)
-\frac{1}{\ln 2}e^{1/\bar{\gamma}_{e,n}}{\rm E_1}\left(\frac{1}{\bar{\gamma}_{e,n}}\right) \right\}^{+}\nonumber
\end{align}
where $\bar{\gamma}_{e,n}=\frac{p_n}{\sigma^2}\mathbb{E}\{\beta_n\}$, and ${\rm E_1}(x)=\int_x^\infty \frac{e^{-t}}{t}dt$.

Note that in our paper $p_{n}$ is assumed to be fixed, the eavesdropper information rate $r_{e,n}$ in both full and statistical CSI cases is independent of $\rho_k$. Therefore, the analysis of both cases is similar. That is, with \eqref{eqn:sr1} substituted by \eqref{eqn:sr2}, the problem (P-PA) for the case of eavesdropper's statistical CSI can be solved by Algorithm 1 or 2.


\section{Numerical Results}

In this section, we present the numerical results to evaluate the performance of the proposed algorithms. In the simulation setup, we consider an OFDMA network with $N=128$ and $K=8$ mobile users who are located in a cell of 10 m with distance to the BS randomly distributed. 
 The eavesdropper is placed exact $10$ m away from the BS station. The small-scale fading is modeled as i.i.d Rayleigh fading over all subcarriers.
In addition, the power is uniformly allocated on each subcarrier, i.e., $p_{n}=P_t/N$, where $P_t$ is the total transmit power of the BS. Let $E_{\rm sum}$ denote the sum power harvested by all users. For all energy receivers in users' terminals, it is assumed that $\zeta=0.4$. The minimum secrecy rate $\bar{C}$ is assumed to be the same for the all users, i.e., $C_k=\bar{C}, \forall k$. For the information receivers in users' terminals, the noise power is assumed to be $\sigma^2=-30$ dBm.

For performance comparison, we also introduce two schemes in  simulation as benchmarks. For the first scheme, denoted as fixed power splitting (FPS), power splitting ratio $\rho_{k}=0.5 ,\forall k$, is fixed for complexity reduction and $\bs X^*$ is obtained according to \eqref{eqn:opxpb2}. For the second scheme, the subcarrier assignment is fixed (FSA), while each $\rho_{k}$  is optimized by the bisection search  according to Algorithm 2. Specifically, each subcarrier is randomly allocated to one user and then we  use the bisection method to find $\bs \rho^*$ achieving all users' required secrecy rate.

\begin{figure}[t]
\begin{centering}
\includegraphics[scale=.65]{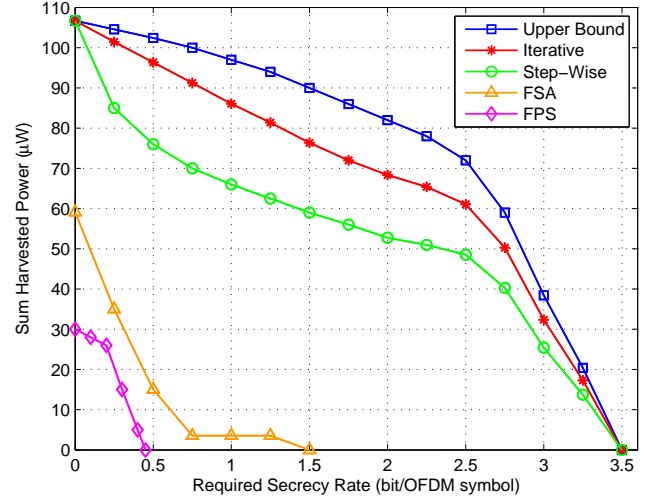}
\vspace{-0.1cm}
 \caption{Achievable $E_{\rm sum}$ versus $\bar{C}$ at total transmit power of $30$ dBm.}\label{fig:f1}
\end{centering}
\vspace{-0.3cm}
\end{figure}

\begin{figure}[t]
\begin{centering}
\includegraphics[scale=.65]{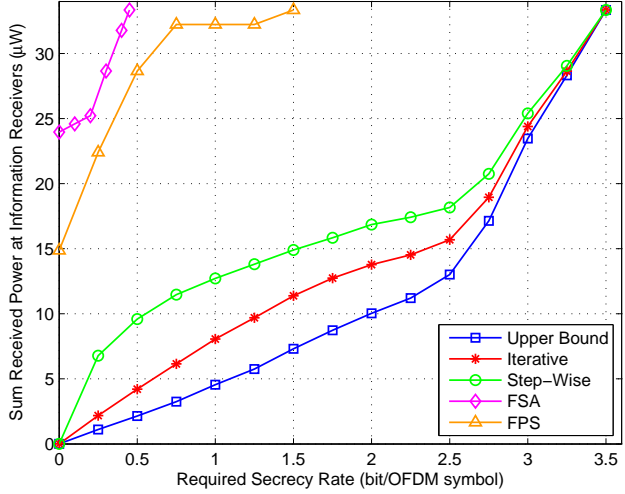}
\vspace{-0.1cm}
 \caption{Power consumption of information receivers versus $\bar{C}$ at total transmit power of $30$ dBm.} \label{fig:f2}
\end{centering}
\vspace{-0.3cm}
\end{figure}

 We first illustrate the achievable harvested power at different required secrecy rate $\bar{C}$ with total transmit power $P_{t}=30$ dBm for full CSI case in Fig. \ref{fig:f1}. It is first observed that for all schemes, $E_{\rm sum}$  decreases with the increase of secrecy rate requirement $\bar{C}$. In addition, $E_{\rm sum}$ falls sharply to zero at $\bar{C}=3.51$ bit/ OFDM symbol. As we have discussed in Section III-B, the optimal $\bs X^*$ for (P-UB) can achieve the same secrecy rate for (P-PA). Therefore, for both the step-wise algorithm and the upper bound, $E_{\rm sum}$ falls to zero at the same $\bar{C}$, where the maximal secrecy rate of both schemes is achieved. It is observed that according to the performance of the upper bound and the iterative algorithm, applying the same power splitting ratio at each user only incurs a little loss in terms of the sum harvested power. Moreover, the proposed step-wise algorithm incurs less than $35\%$ average loss in $E_{\rm sum}$ compared to the iterative algorithm.  Now comparing two proposed algorithms with the benckmarking schemes, both of them show great advantage over FPS and FSA. In addition, the maximal achievable secrecy rate $\bar{C}$ of the FPS and FSA is achieved at around $\bar{C} = 0.45 $ and 1.5 bit/OFDM symbol, respectively, which is much smaller than that of the two proposed algorithms.

Fig. \ref{fig:f2} demonstrates power consumption of information receivers (the sum received power used to satisfy the required secrecy rates) versus different $\bar{C}$ for full CSI case.
 We can observe that with the increase of the required secrecy rate $\bar{C}$, more power should be split into the information receivers for all schemes. Moreover, the proposed step-wise algorithm merely consumes a little more power than the upper bound and the iterative algorithm. In addition, the iterative algorithm performs close to the upper bound. At last, both proposed algorithms consume much less power than FSA and FPS.

 We then illustrate power consumption of information receivers (the sum received power used to satisfy the required secrecy rates) versus the total transmit power $P_{t}$ for full CSI case in Fig. \ref{fig:f3}. It is first observed that with the increase of the transmit power $P_t$,  more power should be split into the information receivers for all schemes. This is because with the increase of the transmit power, the achievable information rate of the eavesdropper $r_{e,n}$ also increases, therefore more power should also be split into the information receivers to guarantee the same secrecy rate. Moreover, the two proposed algorithms consume much less power than FPS and FSA schemes.
\begin{figure}[t]
\begin{centering}
\includegraphics[scale=.65]{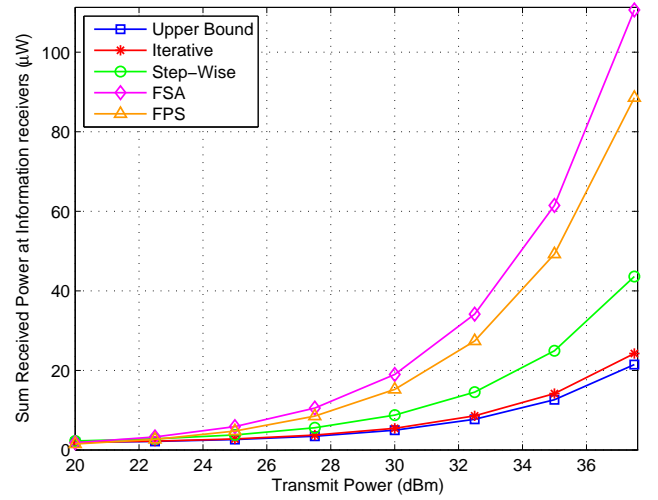}
\vspace{-0.1cm}
 \caption{Power consumption of information receivers versus $P_{t}$ at $\bar{C}=0.5$ bit/OFDM symbol.} \label{fig:f3}
\end{centering}
\vspace{-0.3cm}
\end{figure}

\begin{figure}[t]
\begin{centering}
\includegraphics[scale=.65]{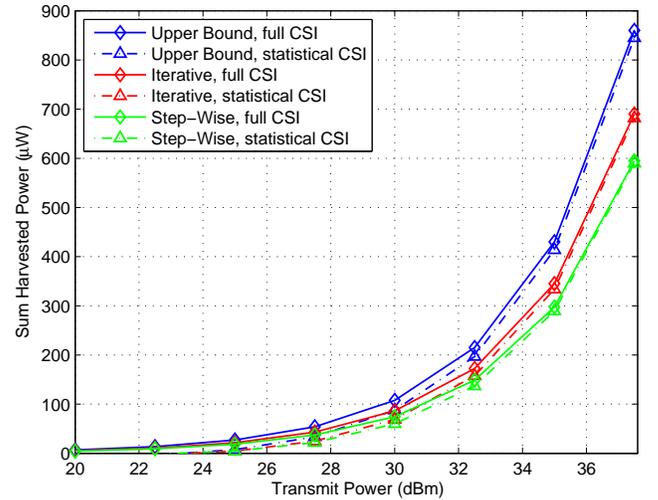}
\vspace{-0.1cm}
 \caption{Achievable $E_{\rm sum}$ versus $P_{t}$ at $\bar{C}=0.5$ bit/OFDM symbol.} \label{fig:f4}
\end{centering}
\vspace{-0.3cm}
\end{figure}

\begin{figure}[t]
\begin{centering}
\includegraphics[scale=.65]{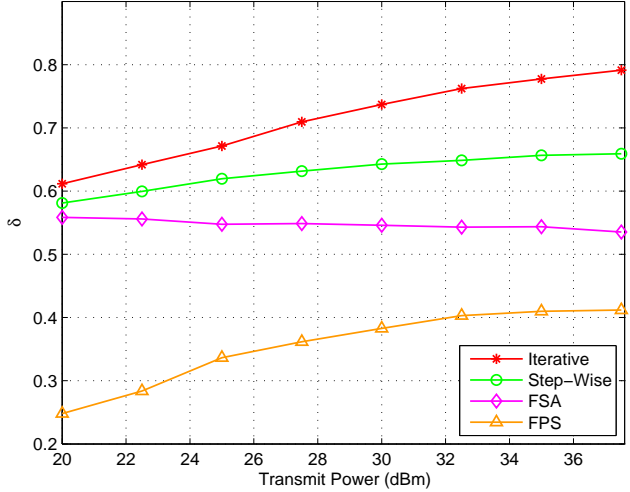}
\vspace{-0.1cm}
 \caption{$\delta$ versus $P_{t}$ at $\bar{C}=0.5$ bit/OFDM symbol. }\label{fig:f5}
\end{centering}
\vspace{-0.3cm}
\end{figure}
In Fig. \ref{fig:f4}, we illustrate the relation between $E_{\rm sum}$ and total transmit power $P_{t}$ with different knowledge of the eavesdropper's CSI (full CSI and statistical CSI) at $\bar{C}=0.5$ bit/OFDM symbol. First, it is observed that the upper bound has the best performance in terms of the sum harvested power. In addition, all the schemes are only achievable when $P_t>20$ dBm and perform very close to each other. Moreover, the upper bound and two proposed algorithms perform better with the full eavesdropper's CSI than only with the statistical CSI. However, with the increase of the transmit power,  each scheme with the statistical CSI performs close to that with full CSI, which coincides with the result in \cite{Ergo2} that additional channel information of the eavesdropper can hardly provide any secrecy rate gain in high SNR region. It is also observed that the iterative algorithm performs close to the upper bound.

We finally demonstrate the relation between $\delta$ and $P_{t}$ in Fig. \ref{fig:f5}, where
$\delta$ is denoted as the ratio of $E_{\rm sum}$ of one specific scheme to that of the upper bound. As it is observed, $\delta$ for proposed iterative algorithm and step-wise algorithm increases with the increase of $P_{t}$, indicating that the two proposed algorithms perform closer to the upper bound with increase of the transmit power $P_t$.  When $P_{t}=37.5$ dBm, $\delta$ of the step-wise algorithm achieves $67\%$, and the iterative algorithm can reach a $\delta=79.5\%$. On the other hand, FPS shows no improvement in terms of $\delta$ as $P_{t}$ increases and it always performs the worst among all schemes.

To conclude the discussion on the above results, the proposed iterative and step-wise schemes greatly outperform FSA and FPS. Specifically, both carefully coordinating subcarrier allocation and selecting power splitting ratio with the adaptation to the channel conditions are insignificant improving the system performance. Furthermore, while the iterative algorithm performs very close to the upper bound, the step-wise algorithm also provides favorable performance, greatly reducing the complexity.

\section{Conclusions}
    This study investigated the joint subcarrier allocation policy and power splitting ratio selection for downlink secure OFDMA-based SWIPT broadband networks. We formulated the problem to maximize the sum harvested power while satisfying the secrecy rate requirements of all users. We studied the performance upper bound and propose two efficient algorithms to tackle the non-convex problems. Numerical results showed that the proposed iterative algorithm performs close to the upper bound and the proposed step-wise algorithm provides a good tradeoff between complexity and performance.


\appendices
\section{Optimal Solution for Performance Upper Bound}
    \begin{algorithm}[tb]
\caption{Dual-Based Method Algorithm for (P-UB)}
\begin{algorithmic}[1]
\STATE \textbf{initialize} $\bs\lambda$.
\REPEAT \STATE Compute
$\rho_{k,n}$  according to \eqref{eqn:oprho} and \eqref{eqn:oprho2}, and then $r_{k,n}^{s}$ according to \eqref{eqn:sr1} or \eqref{eqn:sr2} by replacing $\rho_k$ with $\rho_{k,n}$, for all $k$ and $n$.
\STATE Solve $x_{k,n}$ according to \eqref{eqn:opx} for all $k$ and $n$.
 \STATE
Update $\bs\lambda$ via \eqref{eqn:update} according to the ellipsoid method.
 \UNTIL{$\bs\lambda$ converge.}
\end{algorithmic}

\end{algorithm}
We can derive the Lagrangian function for (P-UB) as follows:
 \begin{align}
 &L(\bs X,\bs \rho ,\bs \lambda )\nonumber\\
 =&\sum_{k=1}^{K} \sum _{n=1}^{N}\zeta \rho _{k,n}p_{n}h_{k,n}+\sum _{k=1}^{K}\lambda _{k}\left (\sum _{n=1}^{N} x_{k,n}r_{k,n}^{s}-C_{k} \right ) \nonumber\\
 =&\sum _{k=1}^{K}\sum _{n=1}^{N}\left (\zeta \rho _{k,n} p_{n} h_{k,n} +x_{k,n}\lambda _{k}r_{k,n}^{s} \right )-\sum _{k=1}^{K}\lambda _{k}C_{k},\label{eqn:Lag}
\end{align}
  where $\bs\lambda =\left [ \lambda _{1},\lambda _{2},...,\lambda _{K} \right ]^{T}$ is the vector of dual variables. The Lagrangian dual function can be obtained as

 \begin{eqnarray}
  g\left ( \bs\lambda  \right )=\max_{\bs X \in \mathcal{T}, \bs\rho \in  \mathcal{R}(\bs X)}L(\bs X,\bs \rho ),
\end{eqnarray}
where $\mathcal{R}(\bs X)$ donate all sets of $\bs \rho$ for given $\bs X$ that satisfy $0\leq\rho_{k,n}\leq1$ when $x_{k,n}=1$ and $\rho_{k,n}=1$ when $x_{k,n}=0$.
We can thus obtain the dual problem as
     \begin{equation}\label{eqn:lamda}
     \min_{\bs \lambda \succeq  0}~g\left ( \bs\lambda  \right ).
     \end{equation}

The dual function $ g\left ( \bs\lambda  \right )$ can be  decomposed into $N$ subproblems which can be solved independently. Each subproblem is obtained as
   \begin{align}
 \max_{\bs X_n \in \mathcal{T}, \bs\rho_n \in  \mathcal{R}(\bs X)} L_{n}(\bs X_{n},\bs \rho_{n})=\zeta p_{n} \sum _{k=1}^{K}  \rho _{k,n}h_{k,n}+\lambda _{k}r_{k,n}^{s} ,\label{eqn:subp}
   \end{align}
and we can rewrite the $L$ in \eqref{eqn:Lag} as
   \begin{align}
L=\sum_{n=1}^N L_n-\sum_{k=1}^{K}\lambda_k C_k.
   \end{align}
\subsection{Optimal Power Splitting Ratio}

   We first seek for the optimal power splitting ratio of each subcarrier. According the
   Karush-Kuhn-Tucker (KKT) conditions \cite{Boyd}, we have
\begin{enumerate}
 \item When $r_{k,n}\geq r_{e,n}$

   \begin{align}
\frac{\partial L_{n}}{\partial {\rho_{k,n}}}=&\sum_{k=1}^{K} \zeta p_{n}h_{k,n}-\frac{\lambda_{k} h_{k,n}p_{n}}{\ln 2\left ( h_{k,n}p_{n}\left ( 1-{\rho _{k,n}} \right)+\sigma ^{2} \right )}\nonumber\\
=&0.
\end{align}
The optimal solution $\rho_{k,n}$ can be readily given by
      \begin{align}
         &&\rho_{k,n}=\left[1-\frac{\lambda _{k}h_{k,n}}{\zeta \ln2 p_{n}\sum_{k=1}^{K}h_{k,n}}+\frac{\sigma ^2}{\ln2 h_{k,n}p_{n}}\right]_0^1,\label{eqn:oprho}\nonumber\\
   \end{align}
   where $[\cdot]^b_a=\max\{\min\{\cdot,b\},a\}$.

 \item When $r_{k,n}< r_{e,n}$

\begin{align}
\frac{\partial L_{n}}{\partial {\rho_{k,n}}}=\sum_{k=1}^{K} \zeta p_{n}h_{k,n}>0.
   \end{align}

The optimal $\rho_{k,n}$ in this case can be obtained as
   \begin{equation}
   \rho_{k,n}=1.
   \end{equation}
\end{enumerate}
  Combining the above two scenarios, the optimal solution $\rho_{k,n}^{*}$ is summarized as
      \begin{eqnarray}
&&\rho_{k,n}^{*}=\begin{cases} \dot{\rho}_{k,n}, ~{\rm if}~r_{k,n}(\dot{\rho}_{k,n})\geq r_{e,n}\\
     1, ~{\rm otherwise},\end{cases}\label{eqn:oprho}
   \end{eqnarray}
where
      \begin{align}
         &&\dot{\rho}_{k,n}=\left[1-\frac{\lambda _{k}h_{k,n}}{\zeta \ln2 p_{n}\sum_{k=1}^{K}h_{k,n}}+\frac{\sigma ^2}{\ln2 h_{k,n}p_{n}}\right]_0^1\label{eqn:oprho2}.
   \end{align}

\subsection{Optimal Subcarrier Assignment}
   Next, substituting $\rho_{k,n}^*$ into $L_{n}(\bs X_{n},\bs \rho_{n})$, the optimal subcarrier assignment policy is given by (the details are easy and omitted here).
   \begin{eqnarray}\label{eqn:opx}
     x_{k,n}^{*}=\begin{cases} 1, ~{\rm if}~k=k^{*}=\arg \max_{k}\mathcal{H}_{k,n}\\
     0, ~{\rm otherwise},\end{cases}
   \end{eqnarray}
where $\mathcal{H}_{k,n}= \zeta p_{n}\sum_{k=1}^{K} \rho _{k,n}^{*}h_{k,n}+\lambda _{k}r_{k,n}^{s} $.

\subsection{Subgradient updating}
  As stated in \cite{Boyd}, the dual problem is always convex and can be solved by using subgradient method. Dual variable $\bs\lambda$ can be updated as follow
    \begin{equation}\label{eqn:update}
\lambda_{k}^{(t+1)}=\left[\lambda_{k}^{(t)}+\alpha_k\left ( C_{k}-\sum_{n=1}^{N}x_{k,n}r_{k,n}^{s}  \right)\right]^+,\forall k.
    \end{equation}

The complexity of this dual based algorithm is analyzed as follows.
For each subcarrier, $ \mathcal{O} ({K}) $ computations are needed. Since the calculation is independent at each subcarrier,
the complexity if $\mathcal{O} ({KN})$ for each iteration. Last, the complexity of subgradient updates is polynomial in $K$ \cite{Boyd}.
Hence, the overall complexity of subgradient method is $\mathcal{O} ({K^{q+1}N})$.
Finally, we present the whole algorithm in Algorithm 3.

\section{Proof for Theorem 1}
In this appendix, we will prove that the optimal subcarriers assignment $\bs X^*$ for (P-UB) is also feasible for (P-PA), given the same inputs (the same channel conditions and secrecy rate requirements).

On one hand, the secrecy rate for user $k$ for (P-UB) is given by
\begin{equation}
     r_{k,ub}^{s}=\sum_{n=1}^{N}x_{k,n}\left [ \log_2\left ( 1+\frac{\left ( 1-\rho _{k,n} \right )p_{n}h_{k,n}}{\sigma^2} \right ) -r_{e,n}\right ]^{+},
\end{equation}
where $r_{e,n}$ is the information rate of the eavesdropper on subcarrier $n$ given in \eqref{eqn:sr1} (full CSI) or \eqref{eqn:sr2} (statistical CSI).

For each user $k$, $r_{k,ub}^{s}$ with the fixed feasible $\bs X$ reaches its maximum when $\rho_{k,n}=0$ for all $n$ that satisfy $x_{k,n}=1$.

On the other hand, for each user $k$, the secrecy rate $r_{k,pa}^{s}$ for (P-PA)  is given by
\begin{equation}
     r_{k,ub}^{s}=\sum_{n=1}^{N}x_{k,n}\left [ \log_2\left ( 1+\frac{\left ( 1-\rho _{k} \right )p_{n}h_{k,n}}{\sigma^2} \right ) -r_{e,n}\right ]^{+},
\end{equation}
which reaches its maximum when $\rho_{k}=0$, and reaches its minimum when $\rho_{k}=1$.

Thus, for the given set of $\bs X$, we obtain that
\begin{align}
\max_{\rho_{k,n}}r_{k,ub}^{s}&=r_{k,ub}^{s}(\rho_{k,n}=0)\nonumber\\
&=\sum_{n=1}^{N}x_{k,n}\left[\log_2\left( 1+\frac{p_{n}h_{k,n}}{\sigma^2} \right) -r_{e,n}\right]^{+}\nonumber\\
&=\max_{\rho_{k}} r_{k,pa}^{s}=r_{k,pa}^{s}(\rho_{k}=0).
\end{align}

In another word, the maximal secrecy rates of both case equal, given set of $\bs X$.

Furthermore, for feasible solution $\bs X^*$ and $\{\rho_{k,n}^*\}$ for (P-UB), we have
   \begin{align}
&0 \leq r_{k,ub}^{s}(\bs X^*,\{\rho_{k,n}^*\})\leq r_{k,ub}^{s}(\bs X^*,\{\rho_{k,n}=0\})\nonumber\\
&=r_{k,pa}^{s}(\bs X^*,\rho_{k}=0).
   \end{align}

Since $r_{k,pa}^{s}$ is a continuous function with respect to $\rho_k$ and monotonically decreasing in $\rho_k$,
there always exists a certain $\rho_k \in \left[0,1\right]$ that satisfies
   \begin{align}
0 \leq r_{k,pa}^{s}(\bs X^*,\rho_{k})=r_{k,ub}^{s}(\bs X^*,\{\rho_{k,n}^*\})\leq r_{k,pa}^{s}(\bs X^*,\rho_{k}=0).
   \end{align}

The proof is thus completed.

\bibliographystyle{IEEEtran}
\bibliography{IEEEabrv,OFDMA}


\end{document}